\documentclass[aps,prl,8pt,twocolumn,unsortedaddress,superscriptaddress]{revtex4}
\usepackage{color}

\usepackage{amsfonts}
\usepackage{amsmath}
\usepackage{amssymb}
\usepackage{bm}
\usepackage[final]{graphicx} 
\begin{document}
\title{Vortex Viscosity in Magnetic Superconductors Due to Radiation of Spin Waves}
\author{A.~Shekhter}
\affiliation{National High Magnetic Field Lab, Tallahassee, Florida 32310, USA}
\author{L.~N.~Bulaevskii} 
\author{C.~D.~Batista}
\affiliation{Los Alamos National Laboratory, Los Alamos, New Mexico 87545, USA}
\begin{abstract}
	In type-II superconductors that contain a lattice of magnetic moments, vortices polarize the magnetic system inducing additional contributions to the vortex mass, vortex viscosity and  vortex-vortex interaction. Extra magnetic viscosity is caused by radiation of spin-waves by a moving vortex. Like in the case of Cherenkov radiation, this effect has a characteristic threshold behavior and the resulting vortex viscosity may be comparable to the well-known Bardeen-Stephen contribution. The threshold behavior leads to an anomaly in the current-voltage characteristics, and a drop in  dissipation for a current interval that is determined by the magnetic excitation spectrum. 
\end{abstract}

\date{\today}
\maketitle

\date{\today}

Magnetic ordering coexists with superconductivity in many different compounds that are known as 
magnetic superconductors. For instance, the  RNi$_2$B$_2$C borocarbides \cite{Canfield98} or 
Ba(Fe$_{1-x}$Co$_x$)$_2$As$_2$ (pnictides)\cite{Chu09}, are superconductors that comprise  a lattice of magnetic moments in their structure.  Coexistence of magnetism and superconductivity is also realized in hybrid systems that are specifically designed for this purpose  \cite{Buzdim04}. A lot of work is being devoted to understanding  the interaction between vortices and magnetic domain walls \cite{domain}, with the main goal of enhancing vortex pinning. However, much less is known about the effect of the magnetic degrees of freedom in the dynamic regimes of flux creep and flux flow.

 Vortices are topological  defects of the superconducting order parameter that emerge  in type II superconductors under the application  of  a strong enough magnetic field. In magnetic superconductors, the magnetic moments couple to the magnetic field of the vortex via Zeeman interaction. Therefore, vortex motion in a magnetic superconductor is accompanied by a dynamic readjustment of the magnetic moments. Since the low-energy excitations of a magnetically ordered system are spin-waves or magnons, a moving vortex can emit magnons under certain conditions. Such magnon emission, in turn, affects the vortex motion.

In this letter we demonstrate that the interaction between vortices and magnetic moments of a type II magnetic superconductor leads to a  contribution to the vortex mass,  vortex viscosity and  vortex-vortex interaction.  We show that a moving vortex can transfer energy to the magnetic moments by emitting spin-waves. This energy transfer gives rise to  a  ``magnetic viscosity''. The inertial part of the response of the magnetic system modifies the dependence of the vortex energy on its velocity. Thus, it generates a magnetic contribution to  the vortex mass.  The  viscosity of this "magnetically dressed'' vortex may also differ significantly from the ``bare'' vortex value. The spin-waves have a well defined dispersion relation $\Omega(\bm{k})$. Similar to the physics of Cherenkov radiation of photons by a moving electron, a superconducting vortex moving with velocity ${\bf v}$ radiates spin-waves with momentum ${\bf k}$ when the kinematic condition $\Omega(\bm{k})={\bf v}\cdot{\bf k}$ is satisfied.  As a consequence, the magnetic viscosity of  a vortex in  an antiferromagnet (or a ferromagnet with easy-axis perpendicular to the applied magnetic field) has a threshold behavior in the vortex velocity, i.e., it is effective only if  $v$ exceeds a critical velocity $v_c$. 

The radiation of photons in the classical Cherenkov problem takes place when 
a charged particle moves faster than the speed of light in a dielectric medium.  In a magnetic superconductor, the threshold velocity $v_c$ depends on the spin-waves velocity, the
size of the vortex core, and the gap of the spin-wave excitation spectrum. 
The  threshold behavior of the magnetic viscosity,  $\eta_m(v)$, can be  observed experimentally as an anomaly in the I-V characteristics of the flux flow regime, near the voltage that corresponds to 
the threshold velocity $v_c$. An important result is that  $\eta_m (v>v_c)$ may be
comparable in magnitude to the core viscosity,  $\eta_c$,  which is induced by excitations of the quasiparticles in the vortex core \cite{Bardeen-Stephen}. Here we only consider temperatures that are well below the magnetic transition temperature $T_N$.


The physics of the magnetic mass is analogous to the polaron  problem.\cite{Feynman} 
The inertia of the magnetic moments adds up to the vortex inertia.    
In this paper we are mainly concerned with the vortex flow regime for which  the effects of 
the vortex mass are not important.\cite{Thouless} 

{\it 1. Spin-vortex interaction in a magnetic superconductor.} 
The magnetic field of a vortex $j$, oriented along the $z$ axis and located at the
spatial coordinate ${\bf R}_j$ in the $xy$ plane, has the form  $\hat{\bf z} H^{z}_v ({\bf r}-{\bf R}_j)$, where  $\hat{\bf z}$ is the unit vector along the $z$-axis. The Fourier components of $H^{z}_v({\bf r})$ are 
$H^{z}_{v{\bf k}}={\phi_0}/(1+\lambda^2{k}^2)$, where $\lambda$ is the London penetration length and $\phi_0$ is the flux quantum.  
In a quasi-static approach ($v$ much smaller than the Fermi velocity), the vortex moving along the trajectory ${\bf R}_j(t)$ 
induces a time-dependent magnetic field $H^{z}_v[{\bf r}-{\bf R}_j(t)]$ that acts on the magnetic subsystem.
Application of a uniform magnetic field induces a uniform internal field $\bm{H}_0={\hat {\bf z}}H_0$ that creates a vortex lattice.  The magnetic field inside the superconductor, ${\bf H}= \hat{{\bf z}} H^{z}({\bf r})$, has the periodicity of the vortex lattice (in an isotropic superconductor the vortices are oriented along the direction of the internal magnetic field.) A moving vortex lattice creates a time-dependent magnetic field ${\bf H} ({\bf r})= \hat{{\bf z}} \sum_j H^{z}_v[{\bf r}-{\bf R}_j(t)]$. For a dense vortex lattice, $H_0 \gg \phi_0 /\lambda^2$,  the periodic component of the magnetic field is small compared to the uniform component,  $h^{z}({\bf r})= H^z({\bf r})-H_0 \ll H_0$.

The Zeeman term is:
\begin{equation}
H_{{\rm int}}=-\sum_{nj}\mu h^z[{\bf r}_{n}-{\bf R}_{j}(t)]S^z_{{n}}(t),
\label{int}\end{equation}
where ${\bf r}_{n}$ are the coordinates of the spins described by the operators ${\bf S}_{n}$ 
and $\mu = g \mu_B$ ($g$ is the gyromagnetic factor and $\mu_B$ is the Bohr magneton). 
We describe spins quantum mechanically, while the vortices are treated as classical objects. 
The vortex-spin interaction given by Eq.~(\ref{int}) leads to the
following equations of motion for spins and vortices 
\begin{eqnarray}
\!\!\! i\hbar \dot{{\bf S}}_{n} \! \! \!&=& \!\![{\bf S}_n, H_M] \! - \! \sum_{{j}}\mu h^z({\bf r}_{n}-{\bf R}_{j})[{\bf S}_{n},S^z_{{n}}]
-i\hbar\Gamma{\bf S}_{n}, \label{spin} \nonumber \\
\\
\eta_{\rm c}\dot{\bf R}_{j}&=&- \frac{1}{L_z}\sum_{n}\mu\frac{\partial h^z[{\bf r}_{n}-{\bf R}_{j}(t)] }{\partial 
{\bf r}_{n}}\langle S^z_{{n}}\rangle + {\bf F}_{\rm L}+{\bf F}_p \nonumber \\
&+& \sum_{\bf j'}{\bf F}_{\rm v}({\bf R}_{j}{\bf R}_{\bf j'}), \label{vortex}
\end{eqnarray}
where $[A,B]=AB-BA$ and $L_z$ is the system length along the $z$ axis, 
$\Gamma$ is the spin relaxation rate
due to coupling of spins  with other degrees of freedom, and 
$\eta_c$ is the vortex Bardeen-Stephen core viscosity.  $F_L$ and $F_p$  are the Lorentz and pinning forces respectively, 
$F_{\rm v}({\bf R}_{{j}},{\bf R}_{{j}'})$ is the non-magnetic force acting 
on vortex ${j}$ due to the presence of other vortices located at ${\bf R}_{{j}'}$. 
These forces are defined per unit of vortex length.
$\langle ...\rangle$ indicates the quantum mechanical average. 
The magnetic system is described by the Hamiltonian, $H_M$, 
which is specified below and accounts for exchange coupling between spins, 
anisotropy, and Zeeman coupling to  ${\bf H}_0$.

The first term in the right hand side of Eq.\eqref{vortex} describes the interaction of a given vortex with other vortices due to polarization of the spin system. When vortices are at rest, this term may lead to a change in the vortex lattice structure depending on the characteristics of the magnetic subsystem and on the applied magnetic field. For dynamical properties, this term leads to additional dissipation due to energy transfer to spin excitations, vortex mass renormalization caused by the inertia of the spins, and change of the moving vortex structure. 
In the following we only consider dynamical effects of the perfect square 
vortex lattice, ${\bf R}_{j}(t)={\bf R}^0_j + \bm{v} t$, with reciprocal 
lattice vectors ${\bf G} = (2\pi/b)(m_x,m_y)$, where $m_{x,y}$ are integers, 
$n_v=1/b^2$ is the vortex density, and $b$ it the vortex lattice parameter. 
We assume that spins are close to equilibrium, i.e., that $\Gamma$ is strong enough in 
comparison with the rate of spin excitations. The  periodic component of the  magnetic field $h^{z}({\bf r},t)$ is weak in comparison with $H_0$. Thus we use the linear response approximation with respect to $h^{z}({\bf r}, t)$ in order to find the Fourier components $\langle S^{z}_{\bf k}(\omega)\rangle$. We note that only nonzero spatial harmonics, ${\bf G}\neq {\bf 0}$, contribute to the coupling of the moving vortex lattice to the magnetic subsystem.  The spatial Fourier components of $h^{z}({\bf r}, t)$ at the initial time $t=0$  are
$h^{z}_{\bf G}(0) = {\phi_0 n_v}/[1+\lambda^2{\bf G}^2]$. We obtain:
\begin{eqnarray}
\mu \langle S^z_{{\bf G}}(t)\rangle &=& \frac{a}{n_v}  h^z_{\bf G}(0) \int_{0}^t  dt'  \chi^{zz}({\bf G},t-t')
e^{-i{\bf G}\cdot{\bf v}t'}, \label{Sz}\\
S^z_{{\bf k}}(t) &=& \frac{a}{L_z} \sum_{n} e^{i{\bf k}\cdot{\bf r}_{n}} S^z_n (t)
\end{eqnarray}
Here $a$ is the lattice parameter of the  magnetic ions and 
$\chi^{zz}({\bf G},t)$ is the longitudinal differential susceptibility of the spin system with respect to the alternating field  $ h^z_{\bf G}(t)$. By expressing $\langle S^{z}(\omega) \rangle$ in Eq.~(\ref{vortex}) via Eq.~(\ref{Sz}),  we obtain the equation of motion for the vortex lattice: 
\begin{eqnarray}
\eta_c\dot{{\bf R}} \!\! \! &=& \! \!\! \frac{-i}{n_v}\sum_{\bf G}  |h^z_{\bf G}(0)|^2 {\bf G}e^{i{\bf G} \cdot {\bf R}(t)}
\!\! \int_{0}^t  \!\! \! dt'   \chi^{zz}({\bf G},t-t')
e^{-i{\bf G}\cdot{\bf v} t'}
\nonumber \\
&+& {\bf F}_{\rm L},
\label{eqm}\end{eqnarray}
where ${\bf R}$ is the center of mass coordinate of the vortex lattice.
Here we neglected the pinning force.
When the vortex lattice moves with constant velocity ${\bf v}$, we obtain from Eq.~(\ref{eqm}):
$\eta_c {\bf v} + {\bf F}_m={\bf F}_{\rm L}$ and
\begin{equation}
{\bf F}_m = - \frac{\bf v}{n_v v^2} \sum_{\bf G}|h^z_{\bf G}(0)|^2   {\bf G}\cdot{\bf v}~
{\rm Im}[\chi^{zz}({\bf G},\omega={\bf G}\cdot{\bf v})].
\label{eq:thetwo}
\end{equation}
Here we introduced the magnetic viscous force ${\bf F}_m$ (per unit of vortex length). 
So far the discussion is general and valid for any magnetic superconductor. 

{\it Magnetic viscous force for an antiferromagnet.}  
We now consider the simple case of a two-sublattice antiferromagnet  with an easy axis anisotropy  along the $x$ direction. 
Thus, we introduce the following magnetic Hamiltonian for spin $S$ moments:
\begin{eqnarray}
H_{M} = J \sum_{\langle l, n \rangle} {\bf S}_l \cdot {\bf S}_n + J_A  \sum_{\langle l, n \rangle} S^x_l S^x_n 
- \mu H_0 \sum_{n}  S^z_n,
\end{eqnarray}
that is defined on a bipartite lattice with coordination number ${\cal Z}$. Here $\langle l,n \rangle$ indicates that $l$ and $n$
are nearest-neighbor (NN) lattice sites, $J$ is the exchange constant between NN, and
$J_A>0$ is the amplitude of the exchange anisotropy term.   
This model is a good starting  point for describing  magnetic 
superconductors that have a Neel  temperature, $T_N$, lower than the superconducting critical temperature $T_c$. While the exchange term stabilizes a two-sublattice antiferromagnetic ordering along the $x$ direction (easy axis), $H_0$ induces a uniform component along the $z$-axis. This canted antiferromagnetic phase persists at $T \ll T_N$ until the field reaches the saturation value $H_s$, i.e., the staggered magnetization component  $\mu\langle S^x_n  \rangle = \pm\mu S\sqrt{1-H_0^2/H_s^2}$ vanishes for  $H_0\geq H_s$ (different signs correspond to different sublattices).

We will compute the longitudinal susceptibility, $\chi^{zz}({\bf k}, \omega)$, in the spin-wave approximation. Since the natural upper cut-off for the momenta of the vortex lattice field is $1/\xi \ll 1/a$, we  compute $\chi^{zz}({\bf k}, \omega)$ in the long wave-length limit. The result is $\chi^{zz}({\bf k},\omega)=0$ for $H \geq H_s$ and 
\begin{align}\label{eq:theone}
\chi^{zz}(\omega,{\bf k}; H_0) 
= \frac{ \Omega_1^2+s_1^2k^2}{\Omega_k^2 - (\omega + i \Gamma)^2} 
\end{align}
for  $H_0 \leq H_s$. The saturation field is $H_s=(\mu/a^3) S/\chi_0$, where $\chi_0=\chi^{zz}(\omega=0,k=0)$ is the static susceptibility. Here $ \Omega_1 = (\mu/\hbar a^{3/2}) S \sqrt{ J_A {\cal Z} (1-H^2_0/H^2_s)}$ and $s_1= (\mu/\hbar a^{1/2}) S \sqrt{J (1-H^2_0/H^2_s)}$. The dispersion $\Omega_k$ of the spin-wave excitations that 
couple to the vortex field is \cite{Akhiezer} 
\begin{align}\label{eq:aaa}
& \Omega_k \approx \sqrt{\Omega_0^2+s_0^2k^2}, \;\;
\Omega_0= \frac{ {\cal Z} S}{\hbar} \sqrt{1 - \frac{H^2_0}{H^2_s}}  \sqrt{J_A (2 J+ J_A)} , \notag\\
& s_0 = \frac{a}{\hbar} \sqrt{ \cal Z } S  \sqrt{ 2 J -  J_A }  \Big[ \frac{J_A}{2} + \left(1 - \frac{H^2_0}{H^2_s}\right) 
\left ( J +  \frac{J_A}{2} \right )  \Big]^{1/2}. 
\end{align}
Here $\Omega_0$ and $s_0$ are the spin wave gap and velocity.  
For  $H_0>H_s$, the staggered magnetization is zero  and the radiation of spin-waves vanishes at low temperatures because the magnetic field cannot change the already saturated amplitude of the uniform magnetization. 

By inserting Eq.~(\ref{eq:theone}) into Eq.~(\ref{eq:thetwo}), we obtain:
\begin{equation}
{\bf F}_m = - \frac{\bf v}{2 n_v v^2} \sum_{\bf G}|h^z_{\bf G}(0)|^2   
\frac{{\bf G}\cdot{\bf v}  ~\Gamma (\Omega_1^2+s_1^2G^2)}{\Omega_G[({\bf G}\cdot{\bf v}-\Omega_G)^2+\Gamma^2]}.
\label{eq:thetwob}
\end{equation}
To proceed, we replace the sum over $\bm{G}$ by an integral by assuming that $\Gamma/s_0 \gg 2\pi/b$, i.e., that the integrand varies over a length scale much longer than $2\pi/b$. When  $\Gamma \ll k^* v $, with $k^*= {\Omega_0}/{\sqrt{v^2-s_0^2}}$, the Lorentzian can be replaced by a delta function, ${\rm Im}[\chi^{zz}({\bf k},\omega={\bf k}\cdot{\bf v})]
\simeq (\Omega_1^2+s_1^2k^2) \delta(\Omega_k-{\bf k}\cdot {\bf v})/\Omega_k$. 
In this form the kinematic condition of the radiation is explicit. 
As we mentioned above, linear response is applicable when the spin relaxation is fast compared to the rate at which magnons are created by the vortex lattice. We can now specify this condition explicitly: $\Gamma \gg a^{3/2} \Omega_1\phi_0 /(2\pi \sqrt{\Omega_0} \lambda)^2$. 
By performing the angular integration, ${\bf k} \cdot {\bf v}=k v\cos{\phi}$, 
we obtain the magnetic viscous force per unit of vortex length for $v>s_0$: 
\begin{eqnarray}\label{eq:integral}
F_m(v) \!\! = \!\! \frac{\pi}{v} \left[\frac{\phi_0}{\lambda^2}\right]^2 \!\!\! \int_{0}^{\frac{2\pi}{\xi}} \!\!\!\! \frac{kdk}{(2\pi)^2} 
\frac{\Omega_1^2+s_1^2k^2}{\sqrt{v^2k^2- \Omega_k^2}} \frac{\theta(k-k^*)}{[k^2+\lambda^{-2}]^2}, 
\end{eqnarray}
where $\theta(x)=1$ if $x>0$ and zero otherwise. 
Unlike the Cherenkov electron, which is a point-like object and generates a field with arbitrary high  wave vector, the vortex has a finite core size and its spatial Fourier components are exponentially suppressed  for  $k\gtrsim2\pi/\xi$. This determines the upper integration limit in  Eq.~(\ref{eq:integral}). 

We will now analyse this equation for $F_m(v)$ in the range: 
\begin{equation}
 \frac{2\pi s_0}{b}, \frac{a^{3/2} \Omega_1}{ \sqrt{\Omega_0}} \left(\frac{\phi_0 }{(2\pi  \lambda)^2}\right) \ll \Gamma \ll \frac{2\pi s_0}{\xi}\,.
\end{equation} 
As the vortex velocity increases, the value of $k^*(v)$ starts at $k^*(s_0)=\infty$, for $v=s_0$, and decreases monotonically for $v>s_0$. Depending on the value of $k^*(v)$  relative to the  upper 
integration limits, Eq.~(\ref{eq:integral}) has two different regimes: 

(i)  $k^*$ is above the upper integration limit,  $k^*>2\pi/\xi$. The magnetic viscous force vanishes, $F_{m} = 0$,  for $v<v_c=[s^2+ (\xi\Omega_0/2\pi)^2]^{1/2}$.

\begin{figure}[ht!]
\centerline{\includegraphics[bb=20 20 300 500,width=0.6\columnwidth]{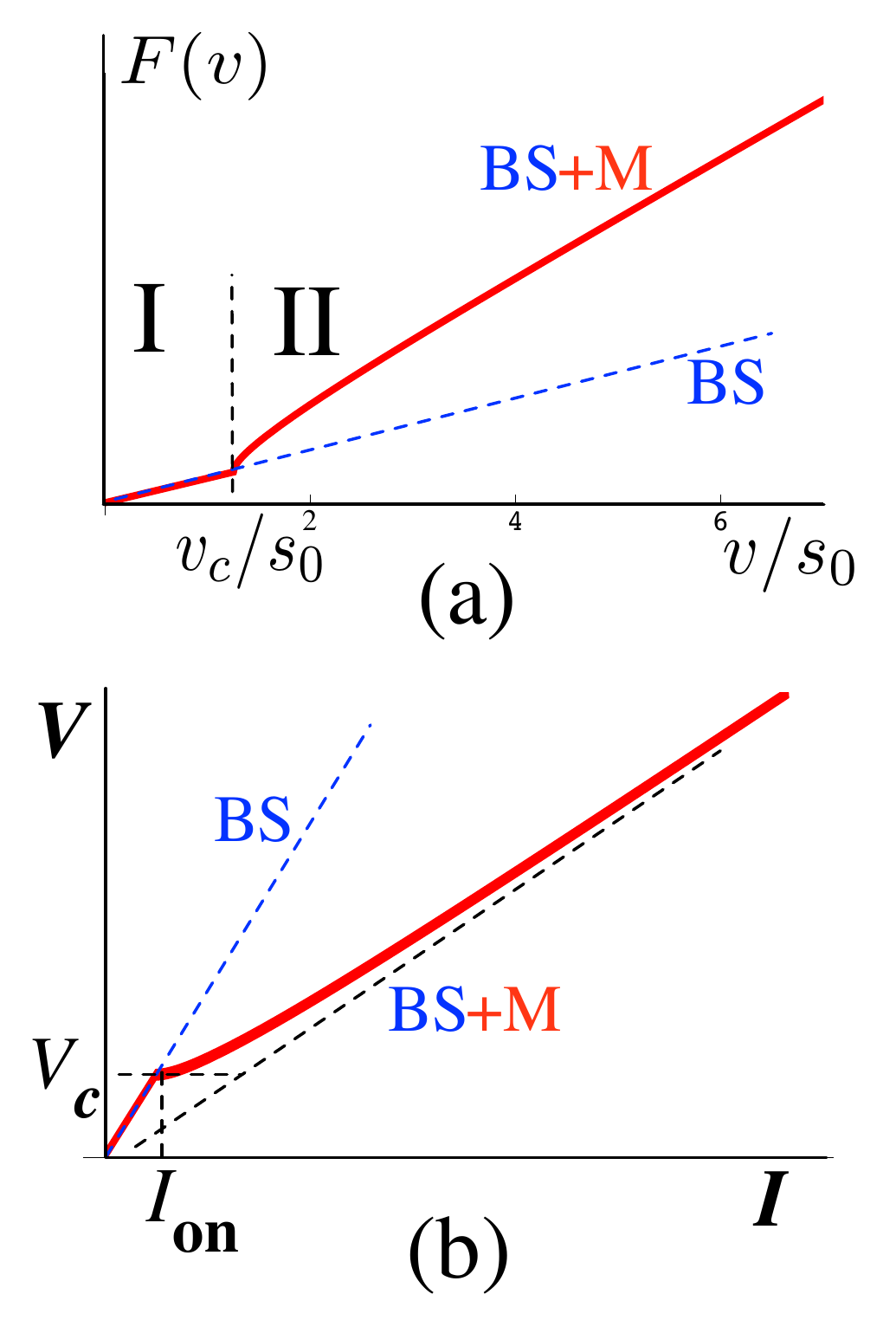}}
\caption{ (Color online) \textbf{(a)}  Total viscous force in arbitrary units as a function of $v/s_0$, as given by Eq.~(\ref{eq:integral}).  The blue dashed line represents the effect of Bardeen-Stephen viscosity due to vortex core alone. \textbf{(b)} Effect of the magnetic viscous force on the I-V characteristics. Magnetic viscous force results in a voltage drop relative to the BS result for currents above $I_{\rm on}$ corresponding to the vortex velocity $v_c$.}
\label{fig:cherenkov}
\end{figure}

(ii) For $k^*<2\pi/\xi$ the integration in Eq.~(\ref{eq:integral}) has the limits $k^*<k<2\pi/\xi$. At $v=v_c$ the dissipation function has a threshold behavior $F_{m} \propto \sqrt{v-v_c}$. 
Note that the threshold is  sharp because $\Gamma \ll 2\pi s_0/\xi$.  As the velocity increases beyond  $[v_c^2 + (\xi\Omega_0/2\pi)^2]^{1/2}$, but remains small enough to ensure $k^*(v)\gg2\pi/b $, the viscous force has an approximately linear dependence on the vortex velocity and we can define a magnetic viscosity via   
\begin{equation}
\label{eq:Pm-linear}
F_{m}(v) =\frac{\eta_m}{v} (v^2-s_0^2+2 s_1^2/\chi_0)\,,  
 \;\; \eta_m = \frac{\chi_0}{16}\left( \frac{\phi_0}{\lambda^2}\right)^2  \frac{1}{\Omega_0}\,.
\end{equation}
Fig.~\ref{fig:cherenkov} shows the dependence of the magnetic viscous force $F_m$ on the vortex velocity. As $H_0$ approaches $H_s$,  the amplitude of the periodic part of the magnetic field decreases and  
the value of $F_{max}$ eventually vanishes. 

As  temperature increases and approaches $T_N$, the spin-wave relaxation rate  $\Gamma$ increases. The dynamics becomes relaxational near the magnetic transition\cite{LandauKhalatnikov,Chaikin}. In this  regime, the magnetic viscosity does not exhibit a threshold behavior in the vortex velocity. 

{\it 3. Experimental signatures.} 
We can estimate the magnitude of the magnetic viscosity by evaluating Eq.~(\ref{eq:Pm-linear}). For  typical values of the relevant parameters, $\mu\sim10\mu_B$, 
$T_c \simeq 10$ K, $H_{c2}\sim10\;\text{T}, \rho_n \sim100\;\mu\Omega \text{cm},\lambda\sim200\;\text{cm}$, $s\sim 50\,\text{m/s}$ 
and $\Omega_0 = 2$ K, we obtain $\eta_m\sim 10^{-7}\; \text{g/(cm $\cdot$ sec)}$. This value is comparable to the BS core viscosity\cite{Bardeen-Stephen} $\eta_c=(H_{c2}\phi_0\sigma_n)/{c^2}\sim 10^{-7}\; \text{g/(cm $\cdot$ sec)}$. Clearly, the magnetic contribution to the vortex viscosity is more important for nearly isotropic magnetic materials (small gap $\Omega_0$) such as superconducting compounds or heterostructures whose magnetic moments are provided by transition metal ions.

We  will now discuss the current-voltage characteristics. 
In the experimental set up,  the voltage $V$  is measured at a fixed current. 
The electric current exerts a Lorentz force, $F_L = (\phi_0/c)I$, on the vortex. The voltage $V$ 
is proportional to the velocity of the vortex flow via the Josephson relation $V=(\hbar/2e) (v/b)$. 
The vortex velocity is determined from the balance between the Lorentz force, $F_L$, and the viscous force 
given by Eq.\eqref{eq:thetwo}.
As the current exceeds the value $I_{\rm on}$ at which the vortex velocity (determined by the Bardeen-Stephen viscosity only) equals to the onset velocity, $v_c$, the character of I-V curve changes. Right above $v_c$ (onset of region II in  Fig.~\ref{fig:cherenkov}), the magnetic contribution to the vortex viscosity is nonzero and increases fast, $F_m \propto \sqrt{v-v_c}$. Consequently, vortex velocity is weakly dependent on $I$ near the threshold and $V-V_c \propto (I - I_{on})^2$ where $V_c = (\hbar/2e)(v_c/b)$. Once the current is large enough to propel vortex to a larger speed where magnetic viscosity saturates to a constant value, the linear behavior, $V \propto I$, is recovered. The slope of linear behavior of $V$ vs $I$ in region II is smaller compared with region I because the total viscosity is larger.

It is important to determine if the critical vortex  velocity $v_c$, necessary to reach 
the onset of magnetic dissipation, is experimentally attainable. 
The limiting factors are heating and  the Larkin-Ovchinnikov 
instability of the vortex flow at high velocities.\cite{Larkin,Huebener} 
A current density of $\sim50\,\text{KA/cm}^2$ is needed to reach vortex 
velocities close to the spin-wave velocity, $s\sim50\,\text{m/s}$. 
If heating is the limiting factor, such currents can be reached with pulsed measurements. In addition, $s$ decreases  for increasing  magnetic field and temperature. It is hard to get a reliable estimate of the critical velocity for the Larkin-Ovchinnikov instability, which is given approximately by $v^*\approx v_F \sqrt{\tau_{\text{el}}/\tau_{\text{in}}}$ ($v_F$ is the Fermi velocity and $\tau_{\text{el,in}}$ are the elastic and inelastic scattering times respectively). 
Measurements of YBCO films yield $v^*\sim1000$ m/s \cite{Huebener}.

In conclusion,  moving vortices radiate spin-waves in superconducting antiferromagnets
for $v>v_c$. This effect decreases the flux flow resistance, i.e., it reduces energy dissipation. 
The onset of magnetic viscosity appears in the I-V characteristic curve 
as a  local deviation at the threshold voltage that 
is determined by a vortex velocity $v_c$.  For materials with $T_N<T_c$, 
we predict a drop in the resistance as  temperature drops below $T_N$. 
The same effect can be observed in hybrid structures made of alternating superconducting 
and magnetic layers for a magnetic field perpendicular to the layers. 
The magnetic  layers can be antiferromagnetic or ferromagnetic with easy axis parallel to the layers.

\begin{acknowledgements}
We thank Leonardo Civale and Boris Maiorov for discussion of experimental details. Research supported by the U.S. Department of Energy, Office of Basic Energy Sciences, Division of Materials Sciences and Engineering. 
\end{acknowledgements}


\begin{thebibliography}{99}

\bibitem{Canfield98} 
P.C.~Canfield, P.L.~Gammel, and D.J.~Bishop, Physics Today ~{\bf 51}, 40 (1998). 

\bibitem{Chu09} J-H.~Chu {\it et al},  Phys.~Rev.~B ~{\bf 79}, 014506 (2009).

\bibitem{Buzdim04} A. Buzdim, Nature Materials ~{\bf 3}, 751 (2004).

\bibitem{domain} D.B. Jan, J.Y. Coulter, B.B. Maranville, M.E. Hawley, L.N. Bulaevskii, M.P. Maley, F. Hellman. 
X.Q. Pan, Q.X. Jia, J. Appl. Phys. {\bf 82}, 778 (2002)L.N. Bulaevskii, E.M. Chudnovsy, M.P. Maley, 
Appl. Phys. Lett. {\bf 76}, 2594 (2000).. 

\bibitem{Ivan}J.I. Martin, M. Velez, J. Ngeues, I. Schuler, Phys. Rev. Lett. {\bf 79}, 1929 (1997); M. Velez, J.I. Martin,
J.E. Villegas, A. Hoffmann, E.M. Gonzales, J.L. Vicent, I. Schuller, J. Magn. Magn. Materials, {\bf 320}, 2547 (2008).

\bibitem{Lange02} M.~Lange, M.J.~Van Bael, V.V.~Moshchalkov, and Y.~Bruynseraede, Appl.~Phys.~Lett. ~{\bf 81}, 322 (2002).

\bibitem{Bardeen-Stephen}J.~Bardeen and M.~Stephen, Phys.~Rev. ~{\bf 140}, A1197 (1965).

\bibitem{LandauKhalatnikov} L.D.~Landau, I.M.~Khalatnikov, Dokl.~Acad. Nauk. SSSR \textbf{96}, 469 (1954);  
\emph{Collected Papers of  L.~D.~Landau}, edited by D.~ter~Haar (Pergamon, London, 1965). 

\bibitem{Feynman} R.P.~Feynman, \emph{Statistical Mechanics}, (Benjamin, Reading, Mass., 1972).

\bibitem{Vinokur} G.~Blatter, M.V.~Feigelman, V.B.~Geshkenbein,	A.I.~Larkin, and V.M.~Vinokur, Rev. Mod. Phys. ~\textbf{66}, 1125
(1994).

\bibitem{Thouless} D. J. Thouless, and J. R. Anglin, Phys.~Rev.~Lett. ~\textbf{99},  105301 (2007). 

\bibitem{LLP} L.~D.~Landau, and E.~M.~Lifshitz, \emph{Course of Theoretical Physics, Vol.~8}, E.~M.~Lifshitz, and L.~P.~Pitaevskii, \emph{Electrodynamics of Continuous Media} (Pergamon, Oxford, 1984).

\bibitem{deGennes} P.~D.~de~Gennes, Superconductivity of Metals and Alloys,  (Benjamin, New York, 1966). 


\bibitem{Akhiezer} A.I.~Akhiezer, V.G.~Bar'yakhtar, and M. I.~Kaganov, Sov.~Phys.~Usp.~\textbf{3}, 567 (1961). [Usp. Fiz. Nauk~\textbf{71}, 533].

\bibitem{Chaikin} P.M.~Chaikin, and T.~Lubensky, \emph{Principles of Condensed Matter Physics},  (Cambridge University Press, Cambridge, 1994). 

\bibitem{Larkin} A.I.~Larkin, and Yu.N.~Ovchinnikov, Zh.~Eksp.~Teor.~Fiz.~\textbf{68}, 1915 (1975)  JETP~\textbf{41}, 960 (1976)]. A.~Schmid, and W.~Hauger, J.~Low~Temp.~Phys. ~\textbf{11}, 667 (1973). 

\bibitem{Huebener} S.G.~Doettinger, R.P.~Huebener, R.~Gerdemann, A.~K\"uhle,
and S. Anders, Phys.~Rev.~Lett. ~\textbf{73}, 1691 (1994).

\end{thebibliography}
\end{document}